# Ultrafast flow of interacting organic polaritons

Giovanni Lerario[1], Dario Ballarini[1*], Antonio Fieramosca[1], Alessandro Cannavale[1,3], Armando Genco[2,3], Federica Mangione[1], Salvatore Gambino[1,3], Lorenzo Dominici[1,2], Milena De Giorgi[1], Giuseppe Gigli[1,3], Daniele Sanvitto[1]

[1] CNR NANOTEC − Institute of Nanotechnology, via Monteroni, 73100, Lecce, Italy

[2] CBN-IIT, Istituto Italiano di Tecnologia, Via Barsanti, 73100, Lecce, Italy

[3] Dipartimento di matematica e fisica "Ennio De Giorgi", Università del Salento, Via Arnesano, 73100 Lecce, Italy

* email: dario.ballarini@nanotec.cnr.it

**The strong-coupling of an excitonic transition with an electromagnetic mode results in composite quasi-particles called exciton-polaritons, which have been shown to combine the best properties of their bare components in semiconductor microcavities. However, the physics and applications of polariton flows in organic materials and at room temperature are still unexplored because of the poor photon confinement in such structures. Here we demonstrate that polaritons formed by the hybridization of organic excitons with a Bloch Surface Wave are able to propagate for hundreds of microns showing remarkable third-order nonlinear interactions upon high injection density. These findings pave the way for the studies of organic nonlinear light-matter fluxes and for a technological promising route of dissipation-less on-chip polariton devices working at room temperature.**

## Introduction

Exciton polaritons are hybrid quasi-particles, arising from the strong coupling between excitons and photons, which possess both the features of their bare components.[1,2,3] Photons, which are massless and non-interacting particles, lighten the exciton mass down of 3 orders of magnitude,[4,5,6] while excitons carry their nonlinear properties, which are 4 orders of magnitude higher than in standard nonlinear optical media.[3,7,8] These attributes led, in the last 10 years, to the observation of fascinating new physics in solid state systems, such as polariton condensation,[9,10,11] superfluidity,[12,13] quantized vortices[14,15] and their non-linear dynamics.[16] Recently, exciton polaritons have also attracted a strong technological interest as ultrafast and dissipative-less optical devices.[17,18,19,20] In this context, exciton polaritons are perfect candidates for this scope: the excitonic component makes them extremely sensitive to small power changes, while the photonic part allows for information manipulation with high speed, efficiency and no heat dissipation.[21,22,23]

The overwhelming majority of the research was developed adopting inorganic-based systems (typically semiconductor quantum wells) working at cryogenic temperatures, which limits their potential technological applications. For their ease of fabrication, low costs and high binding energies,[24,25] organic polaritons are an emerging field with outstanding potentialities, as demonstrated by the recent observation of room temperature condensation.[26,27] Conventional polaritonic structures are planar microcavities, where the optical active layer is embedded between two Distributed Bragg Reflectors (DBR). However, the high dissipation rate of organic planar microcavities (short lifetimes) together with their relatively low group velocity (typically 1-2 µm/ps), prevented so far the observation of the fascinating physics related to flows of light-matter particles as well as any possible applications in cascadable-on-chip technologies, deeply investigated in inorganic systems.[28,29,30,31] Here we show organic polariton propagation, for distances longer than 120 microns and with group velocities of about 50% the speed of light, in a single DBR structure, working at room temperature.

To obtain such fast velocities, an organic exciton is coupled with a high-quality Bloch surface wave (BSW). The BSW is an evanescent, lossless and propagating optical mode that differs from other surface modes like Tamm states or surface plasmons since, although possessing a very high group velocity,[32,33,34] it does not undergo dissipation caused by metallic losses. Due to their high quality factor, BSWs are widely studied for sensing application, and their remarkable potentialities in the field of on-chip optical manipulation have been explored only recently.[35,36,37,38] Differently from bare optical BSW investigated in these works, the excitonic component of the BSWPs allows the exploitation of non-linear polariton interactions. Here we demonstrate the blue-shift of the polariton resonance with increasing particle densities, enabling the local dynamic manipulation of the BSWP dispersion by external laser pulses, and also laying the groundwork for polariton devices with logic functionalities operating at room temperature.

**Materials and Methods**

The DBR fabrication process consisted in the deposition of seven $TiO_2$ / $SiO_2$ pairs ($d_{TiO2}$= 85nm and $d_{SiO2}$= 120nm) on 130 μm thick glass substrates using an e-beam evaporator.
A 35 nm thick layer of a perylene derivative (Lumogen Red F305), a small molecule with the absorption and emission spectra reported in Fig. 1a, is thermally evaporated onto the $SiO_2$ top layer of the DBR, with a base pressure of around $10^{-7}$ mbar, and at a deposition rate of about 1.0 Å/s.

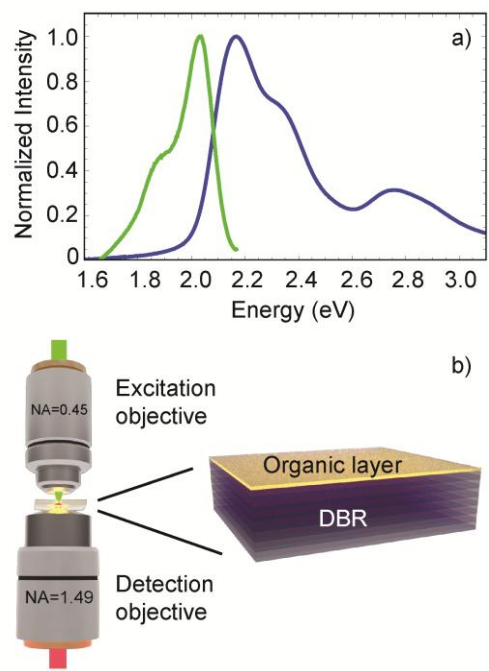

**Figure 1 Material, sample structure and optical setup. a**, Absorption and emission spectra of Lumogen Red F305 in solid state (35nm thermally evaporated thin film). **b**, Illustration of the sample structure and of the leakage radiation microscope setup.

In order to study the Bloch surface wave polariton (BSWP) properties, a leakage radiation microscope (Fig. 1b) coupled to a spectrometer and to a CCD detector is used to measure the energy-resolved, spatial- and momentum-distribution of the signal. The high numerical aperture

(N.A.=1.49) of the oil immersion microscope objective allows the excitation and detection of modes laying beyond the critical angle when positioned at the glass substrate side.

The non resonant BSWP dispersion and spatial distribution is revealed using a continuous wave diode laser at 532 nm focused at the organic deposition side (2 μm spot radius). The laser intensity on the sample surface is limited to 10 W/cm$^2$, preventing organic layer damages.

Resonant measurements are performed by using an energy-tunable 100 fs pulsed-laser. Because the BSWP is an evanescent mode, the incoming pump beam can resonantly inject polaritons only from the substrate side and, therefore, the oil immersion objective is used both for the BSWP excitation and detection, in reflectance configuration. The excitation beam energy and momentum is adjusted to match the BSWP dispersion.

A 300 mm spectrometer with a 1200 gg/mm grating blazed at 600 nm and a high sensitivity CCD camera were used for collecting energy resolved images with an overall resolution of 0.4 nm.

The experimental data are backed up by transfer matrix (TMM) calculations; thicknesses and refractive indexes of the layers constituting the DBR and of the organic material were evaluated by ellipsometric measurements (see Supporting Information).

**Results and Discussion**

The BSWP emission dispersion is shown in Fig. 2a; TMM calculations are superimposed on the same plot. The polaritonic mode (orange line) diverges from the bare optical BSW (blue line) showing the anticrossing behavior typical of strongly coupled systems. Short-propagating evanescent modes associated to the limit of the DBR optical band-gap (sideband modes) are also visible in Fig. 2a at low energies and high momenta.

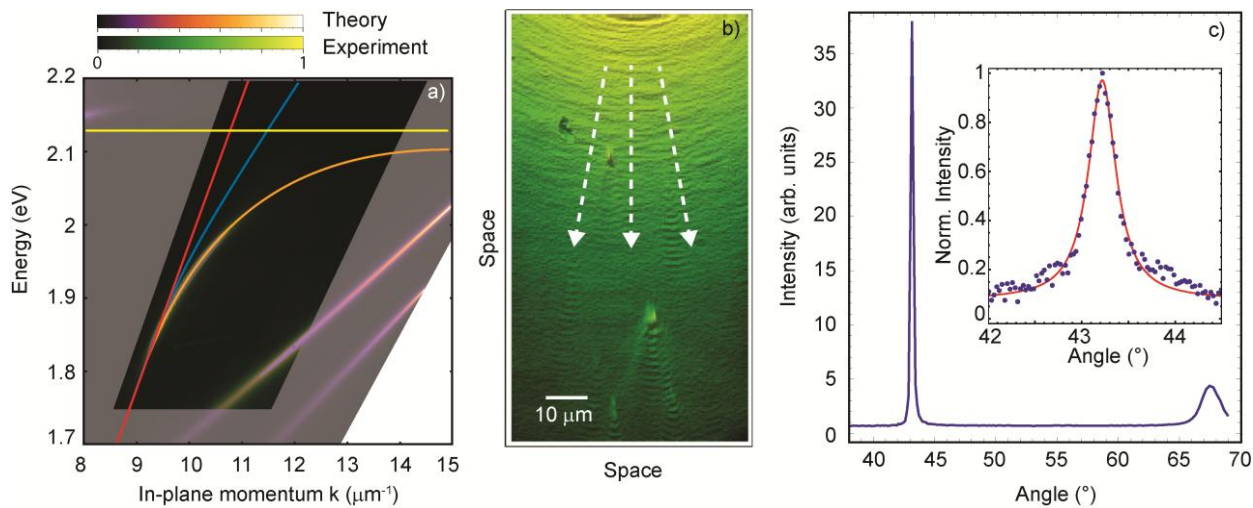

**Figure 2 Dispersion and space propagation of the BSWP. a,** Saturated experimental emission (black background color) superimposed to calculated dispersions (grey background color). The light cone is delimited by the red line, while the blue line indicates the bare optical BSW and the yellow line the exciton energy (2.13 eV). The theoretical BSWP (orange line), obtained with all the parameters extracted from the materials constituting the as-grown structure (see Information), fits perfectly well the experimental results. **b,** Space map of the non-resonant excited polariton emission; white arrows indicate the direction of the polariton flux. **c,** The emission intensity profile at 1.925 eV as a function of the detection angle. In the inset the experimental data (blue dots) are fitted with a Lorentzian function with FWHM of $0.4^o$ (red line).

The BSWP is highly sensitive to small changes of the optical index at the surface of the device,[39] therefore any spatial inhomogeneity of the sample deposition results in a broader BSWP dispersion.[34] Therefore, the sharpness of the polariton mode, clearly visible in the emission profile reported in Fig. 2c, is a first hint of the high local homogeneity of the organic deposition.

Propagation of the polariton fluid on the sample surface is clearly visible in Fig. 2b where a few defects in the structure act as small perturbation on the flow trajectory. This demonstrates also that BSWP are an ideal workbench to study transport properties of dipole-like excitations strongly coupled with an electromagnetic mode, currently at the center of intense investigation and recently proposed to be strongly enhanced in the case of molecules and atoms placed inside an optical cavity.[40,41,42,43]

The polariton flow along a given direction, obtained filtering in momentum space with an angular aperture of about 10 degrees, is shown in Fig. 3a. The energy-resolved emission spectrum of the polariton flow is shown in Fig. 3b. Some "residual" emission, which does not propagate, is detectable close to the excitation beam position (red dashed line), which is related to uncoupled excitons.

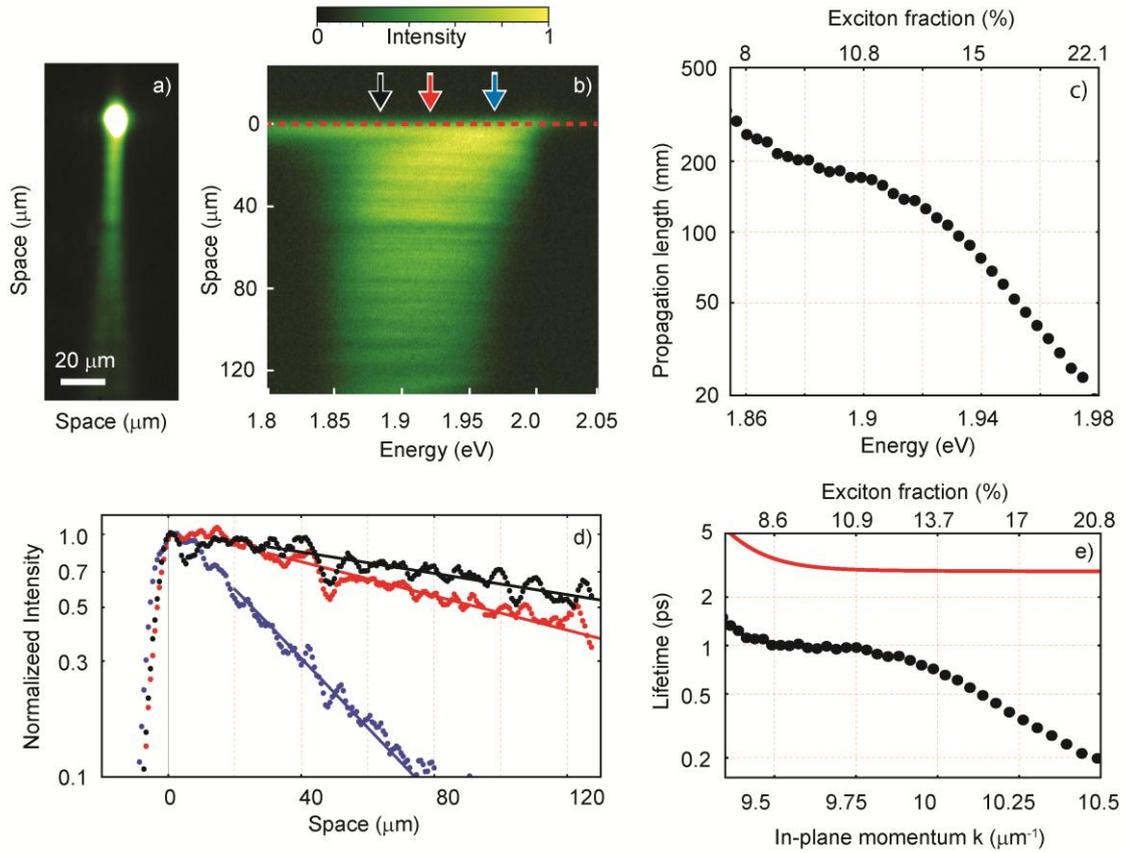

**Figure 3 Polariton propagation and relaxation**. **a**, Space propagation of the BSWP after filtering in momentum-space. **b**, Space propagation in the energy domain. The red dashed line indicates the position of the excitation spot. **c**, Propagation lengths versus energy and exciton fraction. **d**, Intensity profiles (logarithmic scale) of the energy-resolved polariton propagation at different exciton fractions of the polariton mode. At 1.97 eV (blue line), with the excitonic component > 20% and the propagation distance is around 30 μm; at 1.92 eV (red line), the exciton fraction is 12.7% and propagation lengths is 120 nm. At lower energies (1.88 eV, black line), an initial rise of the intensity is observed, followed by a long decay of 200 μm. **e**, Polariton lifetime, versus in-plane wavevector and exciton fraction, evaluated from propagation lengths and group velocity. The red line represents TMM calculation of the bare BSW lifetimes.

Sideband modes and short-living polaritons at higher energies, populated through non-resonant pumping, can also decay into the lower polariton branch but due to their weak-propagating nature they can only affect polariton formation very close to the excitation spot (less than 20 μm).[44,45,46] The amount of this "filling effect" is small, but produces the initial rise of the signal at lower energies as shown in Fig. 3b and Fig. 3d (black arrow). At longer distances, the emission is attributed only to the long propagating BSWP. We can extract the propagation length for different excitonic fractions of the polariton state obtaining values up to 300 μm (8 % of exciton content), with a mean propagation length of about 120 μm (Fig. 3c).

From the experimental propagation length and group velocity ($v_g$), we can obtain the polariton lifetimes ($\tau_{pol}$).[47] The flow speed as a function of energy (E) can directly be extracted from the BSWP dispersion as $v_g = \frac{1}{\hbar}\frac{\partial E}{\partial k}$. Values ranging from 120 μm/ps to 250 μm/ps (see Supporting Information) are obtained, which are about 100 times higher than in standard planar microcavities. At the energy of maximum emission intensity, the polariton lifetime is about 1 ps (see data in Fig. 3e) and the dissipative character of the BSWP is attributed mainly to the excitonic component. Indeed, the DBR mode quality (evaluated from ellipsometric data of the DBR structure) is estimated to be close to E/ΔE = 7000, hence giving a corresponding photon lifetime of about 3 ps (red line in Fig. 3e). This value, although about one order of magnitude higher than in state-of-the-art organic microcavities, is not as high as those reported for inorganic planar microcavities, with lifetimes of some hundreds of ps.[47,48] However, thanks to the high speed of BSWP, the propagation lengths obtained in the present structure are comparable to the ones achieved with the best inorganic-based planar microcavities.[48,49]

To fully exploit the potentiality of polariton propagation in the plane of the device, resonant injection with desired velocity and direction is performed. Resonant excitation below the absorption energy of the exciton is a favorable configuration also to observe the blueshift of the BSWP induced by polariton-polariton interactions only, avoiding the formation of a large reservoir of uncoupled excitons and reducing the heating of the sample and its degradation.

Restricting the field of view to the laser spot region in reflectance configuration, the BSWP appears as a dip in the energy-momentum map, as shown in Fig. 4a for a low energy density signal of 150 μJ/cm$^2$ (as measured before the pump enters the microscope objective).

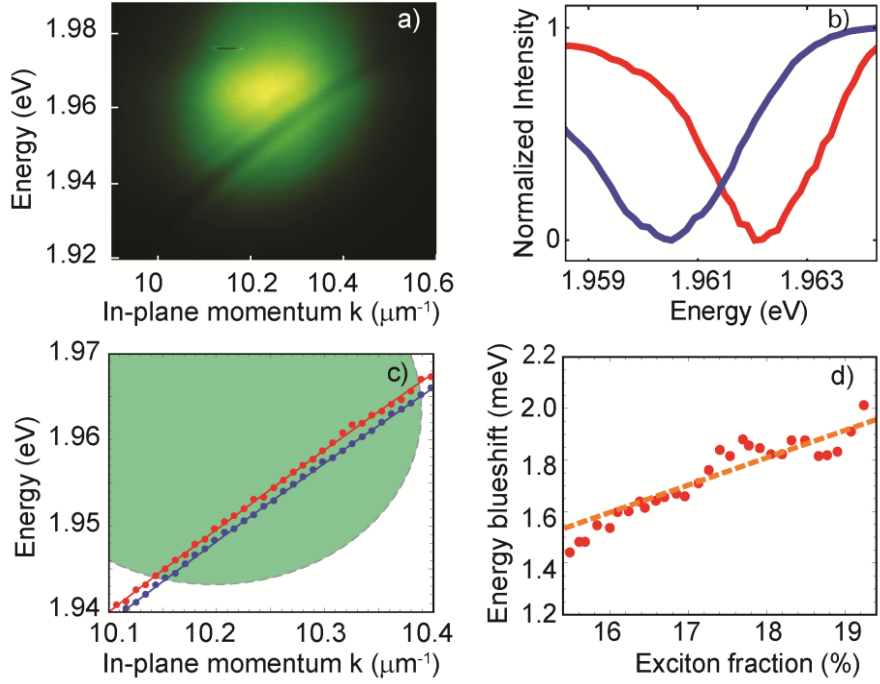

**Figure 4 BSWP non linearities. a,** Bare experimental data of the BSWP dispersion in reflectance configuration at 150 µJ/cm² pumping energy. **b,** Energy-resolved signal at in-plane wavevector k=10.34 µm⁻¹ under 150 µJ/cm² and 10 mJ/cm² excitation energy densities (blue and red lines, respectively), showing the blueshift of the BSWP resonance. **c,** Dots are the experimental BSWP dispersions with 150 µJ/cm² (blue) and 10 mJ/cm² (red) resonant pump pulse. TMM calculations (continuous lines) of the BSWP dispersion perfectly match the experimental ones when considering an exciton energy of 2.13 eV (blue) and a blue-shifted (8 meV) exciton (red). The green ellipse depicts the extension (FWHM) of the laser spot in energy-momentum space. **d,** The expected blueshift as a function of the excitonic fraction (orange line) fits the experimental results (red dots) for an exciton blueshift of 8meV.

When the pumping energy density is increased to $\rho_E$=10 mJ/cm², the BSWP mode blue-shifts in energy. The blue (red) dots in Fig. 4c are the experimental peaks at different energies for the low (high) pumping power. As can be seen, the experimental dispersion at low power perfectly matches the TMM calculation with the exciton peak at 2.13 meV (blue line in Fig. 4c), while, at $\rho_E$=10 mJ/cm², the TMM reproduces the experimental dispersion resulting from an exciton energy blue-shifted of about 8 meV (red line in Fig. 4c). Despite the reflectance spectra are inevitably broader than the actual polariton dispersion due to the setup resolution and the high curvature of the BSWP dispersion that affect the momentum resolution, the shift of the polariton energy is larger than the mode linewidth and can be clearly resolved in our experiments as shown in Fig. 4b. This demonstrate the possibility to use polariton nonlinearities to tune a mode in and out of resonance with an external optical beam.

The wide range of energies covered by the laser pulse allow to observe a trend of increasing polariton blueshift for energies approaching the exciton resonance (Fig. 4d), as expected when increasing the exciton fraction of the BSWP mode.

It is reasonable to consider, from the BSWP depth in the energy-momentum map, that a fraction of about 1% of the pumping power is transferred to the BSWP mode. From the shift of the exciton transition, we can, therefore, estimate the interaction constant which is found in the range of $10^{-3}$ µeV·µm$^2$, in accordance with the one indirectly extracted via non-resonant excitation of organic polaritons in a planar microcavity.[26] The strong coupling with organic molecules thus contributes to enhance the nonlinear behavior of the BSW, however the nonlinearities offered by inorganic polaritons, although being at cryogenic temperatures, are still between 2 to 3 orders of magnitude higher.[13,50]

## Conclusions

Room-temperature ballistic propagation of light-matter excitations in an organic semiconductor, for distances well beyond one hundred of microns, is demonstrated thanks to the adoption of a Bloch surface wave mode. Indeed the BSWP speed exceeds 150 µm/ps, which is two orders of magnitude higher than typical polariton velocities in standard planar microcavities. Moreover, through the resonant injection of a travelling polariton wave-packet, the exciton component of the BSWP manifests in the density-dependent self-energy of the system, showing blue shifts of the polariton resonance for increasing pump powers. This is the first direct measurement of such term, arising from the excitonic components of polaritons, that allows the tuning of the polariton energy and results in a fundamental element for the implementation of polariton on-chip devices.

Although the interaction constant reported in inorganic microcavities are higher than those observed in this work with organic semiconductors, these results are, however, very promising and can be further improved with higher confinement of the electromagnetic field. In this respect, one of the most promising aspects of BSWP is that the surface can be easily patterned to localize the field and enhance the nonlinear optical response of organic materials, paving the way for the realization of ultra fast, and low-loss polariton devices operating at room temperature.

## Acknowledgements

G. L. is grateful to Gianluca Latini for the encouragement at the initial stage of his research path. This work has been funded by the MIUR project Beyond Nano and the ERC project POLAFLOW (Grant N. 308136).

**References**

1. Kavokin, A., Baumberg, J. J., Malpuech, G. & Laussy, F. P. *Microcavities*. (Series on Semiconductor Science and Technology 16, 2007).

2. Bramati, A., Modugno, M. *Physics of Quantum Fluids - New Trends and Hot Topics in Atomic and Polariton Condensates*. (Springer Series in Solid-State Sciences 177, 2013)

3. Sanvitto, D., Timofeev, V. *Exciton Polaritons in Microcavities - New Frontiers*. (Springer Series in Solid-State Sciences 172, 2012)

4. Weisbuch, C., Nishioka, M., Ishikawa, A. & Arakawa, Y. Observation of the coupled exciton-photon mode splitting in a semiconductor quantum microcavity. *Phys. Rev. Lett.* **69,** 3314–3317 (1992).

5. Whittaker, D. *et al.* Motional Narrowing in Semiconductor Microcavities. *Phys. Rev. Lett.* **77,** 4792–4795 (1996).

6. Khitrova, G., Gibbs, H. M., Kira, M., Koch, S. W. & Scherer, A. Vacuum Rabi splitting in semiconductors. *Nat. Phys.* **2,** 81–90 (2006).

7. Tassone, F. & Yamamoto, Y. Exciton-exciton scattering dynamics in a semiconductor microcavity and stimulated scattering into polaritons. *Phys. Rev. B* **59,** 10830–10842 (1999).

8. Savvidis, P. *et al.* Angle-Resonant Stimulated Polariton Amplifier. *Phys. Rev. Lett.* **84,** 1547–1550 (2000).

9. Deng, H., Weihs, G., Santori, C., Bloch, J. & Yamamoto, Y. Condensation of Semiconductor Microcavity Exciton Polaritons. *Science* **298,** 199–202 (2002).

10. Kasprzak, J. *et al.* Bose–Einstein condensation of exciton polaritons. *Nature* **443,** 409–414 (2006).

11. Balili, R., Hartwell, V., Snoke, D., Pfeiffer, L. & West, K. Bose-Einstein Condensation of Microcavity Polaritons in a Trap. *Science* **316,** 1007–1010 (2007).

12. Amo, A. *et al.* Collective fluid dynamics of a polariton condensate in a semiconductor microcavity. *Nature* **457,** 291–295 (2009).

13. Amo, A. *et al.* Superfluidity of polaritons in semiconductor microcavities. *Nat. Phys.* **5,** 805–810 (2009).

14. Lagoudakis, K. G. *et al.* Quantized vortices in an exciton–polariton condensate. *Nat. Phys.* **4,** 706–710 (2008).

15. Lagoudakis, K. G. *et al.* Observation of Half-Quantum Vortices in an Exciton-Polariton Condensate. *Science* **326,** 974–976 (2009).

16. Sanvitto, D. *et al.* Persistent currents and quantized vortices in a polariton superfluid. *Nat. Phys.* **6,** 527–533 (2010).

17. De Giorgi, M. *et al.* Control and Ultrafast Dynamics of a Two-Fluid Polariton Switch. *Phys. Rev. Lett.* **109,** 266407 (2012).

18. Marsault, F. *et al.* Realization of an all optical exciton-polariton router. *Appl. Phys. Lett.* **107,** 201115 (2015).

19. Cerna, R. *et al.* Ultrafast tristable spin memory of a coherent polariton gas. *Nat. Commun.* **4,** 2008 (2013).

20. Sich, M. *et al.* Observation of bright polariton solitons in a semiconductor microcavity. *Nat. Photonics* **6,** 50–55 (2012).

21. Snoke, D. Microcavity polaritons: A new type of light switch. *Nat. Nanotechnol.* **8,** 393–395 (2013).

22. Liew, T. C. H., Shelykh, I. A. & Malpuech, G. Polaritonic devices. *Phys. E Low-Dimens. Syst. Nanostructures* **43,** 1543–1568 (2011).

23. Miller, D. A. B. Are optical transistors the logical next step? *Nat. Photonics* **4,** 3–5 (2010).

24. Lidzey, D. G. *et al.* Strong exciton–photon coupling in an organic semiconductor microcavity. *Nature* **395,** 53–55 (1998).

25. Agranovich, V., Litinskaia, M. & Lidzey, D. Cavity polaritons in microcavities containing disordered organic semiconductors. *Phys. Rev. B* **67,** 085311 (2003).

26. Daskalakis, K. S., Maier, S. A., Murray, R. & Kéna-Cohen, S. Nonlinear interactions in an organic polariton condensate. *Nat. Mater.* **13,** 271–278 (2014).

27. Plumhof, J. D., Stöferle, T., Mai, L., Scherf, U. & Mahrt, R. F. Room-temperature Bose–Einstein condensation of cavity exciton–polaritons in a polymer. *Nat. Mater.* **13,** 247–252 (2014).

28. Ballarini, D. *et al.* All-optical polariton transistor. *Nat. Commun.* **4,** 1778 (2013).

29. Antón, C. *et al.* Dynamics of a polariton condensate transistor switch. *Appl. Phys. Lett.* **101,** 261116 (2012).

30. Liew, T. *et al.* Exciton-polariton integrated circuits. *Phys. Rev. B* **82,** 033302 (2010).

31. Nguyen, H. *et al.* Realization of a Double-Barrier Resonant Tunneling Diode for Cavity Polaritons. *Phys. Rev. Lett.* **110,** 236601 (2013).

32. Descrovi, E. *et al.* Near-field imaging of Bloch surface waves on silicon nitride one-dimensional photonic crystals. *Opt. Express* **16,** 5453–5464 (2008).

33. Descrovi, E. *et al.* Guided Bloch Surface Waves on Ultrathin Polymeric Ridges. *Nano Lett.* **10,** 2087–2091 (2010).

34. Lerario, G. *et al.* Room temperature Bloch surface wave polaritons. *Opt. Lett.* **39,** 2068 (2014).

35. Yu, L. *et al.* Manipulating Bloch surface waves in 2D: a platform concept-based flat lens. *Light Sci. Appl.* **3,** e124 (2014).

36. Angelini, A. *et al.* In-plane 2D focusing of surface waves by ultrathin refractive structures. *Opt. Lett.* **39,** 6391 (2014).

37. Angelini, A. *et al.* Focusing and Extraction of Light mediated by Bloch Surface Waves. *Sci. Rep.* **4,** (2014).

38. Yu, L., Barakat, E., Di Francesco, J. & Herzig, H. P. Two-dimensional polymer grating and prism on Bloch surface waves platform. *Opt. Express* **23,** 31640 (2015).

39. Sinibaldi, A. *et al.* A full ellipsometric approach to optical sensing with Bloch surface waves on photonic crystals. *Opt. Express* **21,** 23331 (2013).

40. Feist, J. & Garcia-Vidal, F. J. Extraordinary Exciton Conductance Induced by Strong Coupling. *Phys. Rev. Lett.* **114,** 196402 (2015).

41. Schachenmayer, J., Genes, C., Tignone, E. & Pupillo, G. Cavity-Enhanced Transport of Excitons. *Phys. Rev. Lett.* **114,** 196403 (2015).

42. Leggio, B., Messina, R. & Antezza, M. Thermally activated nonlocal amplification in quantum energy transport. *EPL Europhys. Lett.* **110,** 40002 (2015).

43. Orgiu, E. *et al.* Conductivity in organic semiconductors hybridized with the vacuum field. *Nat. Mater.* **14**, 1123–1129 (2015).

44. Ballarini, D. *et al.* Polariton-Induced Enhanced Emission from an Organic Dye under the Strong Coupling Regime. *Adv. Opt. Mater.* **2,** 1076–1081 (2014).

45. Coles, D. M. *et al.* Vibrationally Assisted Polariton-Relaxation Processes in Strongly Coupled Organic-Semiconductor Microcavities. *Adv. Funct. Mater.* **21,** 3691–3696 (2011).

46. Michetti, P. & La Rocca, G. C. Exciton-phonon scattering and photoexcitation dynamics in $J$-aggregate microcavities. *Phys. Rev. B* **79,** 035325 (2009).

47. Steger, M., Gautham, C., Snoke, D. W., Pfeiffer, L. & West, K. Slow reflection and two-photon generation of microcavity exciton–polaritons. *Optica* **2,** 1 (2015).

48. Nelsen, B. *et al.* Dissipationless Flow and Sharp Threshold of a Polariton Condensate with Long Lifetime. *Phys. Rev. X* **3,** 041015 (2013).

49. Wertz, E. *et al.* Spontaneous formation and optical manipulation of extended polariton condensates. *Nat. Phys.* **6,** 860–864 (2010).

50. Vladimirova, M. *et al.* Polariton-polariton interaction constants in microcavities. *Phys. Rev. B* **82,** 075301 (2010).

# Ultrafast flow of interacting organic polaritons

# Supporting Information

Giovanni Lerario[1], Dario Ballarini[1*], Antonio Fieramosca[1], Alessandro Cannavale[1,3], Armando Genco[2,3], Federica Mangione[1], Salvatore Gambino[1,3], Lorenzo Dominici[1,2], Milena De Giorgi[1], Giuseppe Gigli[1,3], Daniele Sanvitto[1]

[1] CNR NANOTEC − Institute of Nanotechnology, via Monteroni, 73100, Lecce, Italy
[2] CBN-IIT, Istituto Italiano di Tecnologia, Via Barsanti, 73100, Lecce, Italy
[3] Dipartimento di matematica e fisica "Ennio De Giorgi", Università del Salento, Via Arnesano, 73100 Lecce, Italy

* email: dario.ballarini@nanotec.cnr.it

## Absorption of the Lumogen Red F305

The Napierian absorption coefficient (in cm$^{-1}$) of the Lumogen Red F305, evaluated from 35 nm thick deposition on glass substrate, as a function of the wavelength is shown in Fig. S1. From the absorption coefficient we extract the oscillator strength value inserted in the TMM.

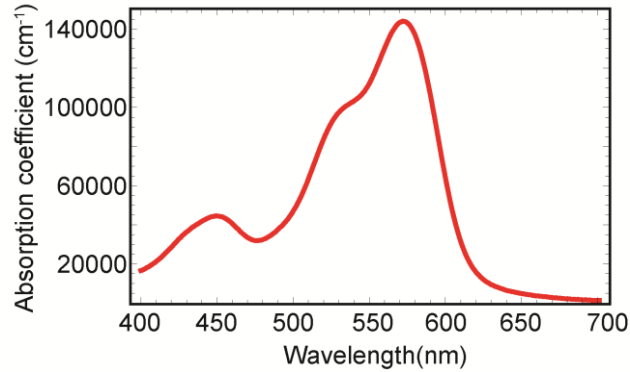

**Figure S1 Lumogen absorption coefficient.** Napierian absorption coefficient spectrum of Lumogen Red F305 in solid state.

## Ellipsometric measurements

Optical indexes and thickness of the layers constituting the DBR are evaluated by preliminary ellipsometric measurements (reported below). The depositions were made on Si substrates in order to have a good refractive contrast between the substrate and the deposited material. The implemented models perfectly match the experimental ellipsometric data; the trustworthiness of the models was checked performing incidence angle scanning (see Fig. S2). The extracted values of the thicknesses–together with profilometer analysis–were also used for the calibration of the DBR fabrication system.

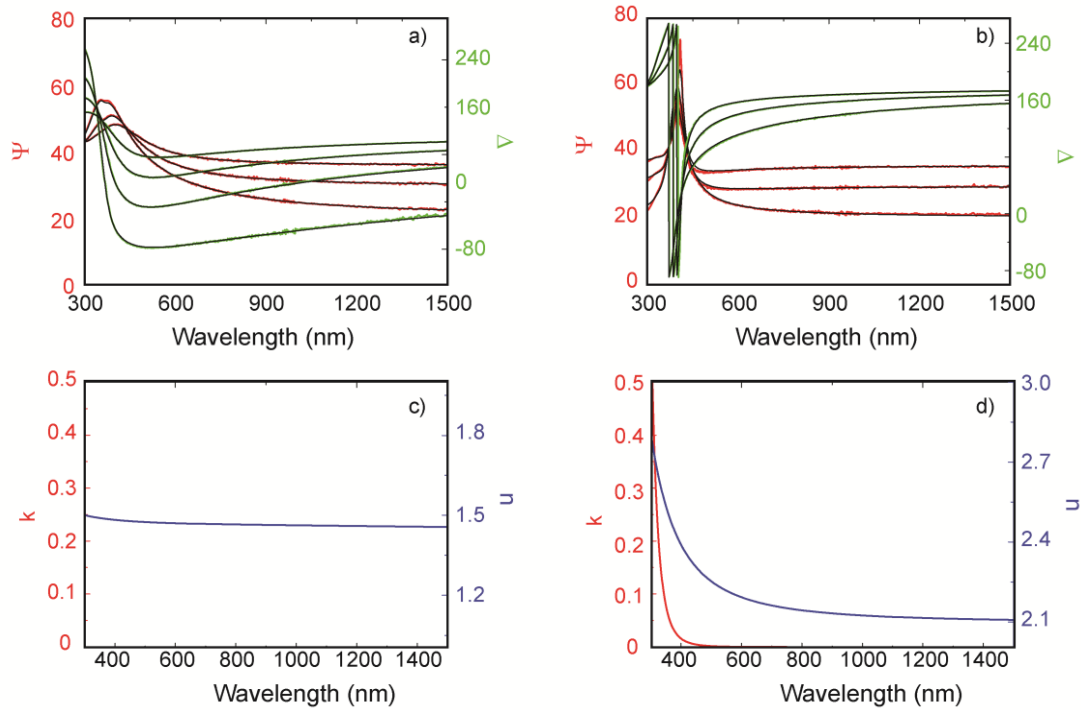

**Figure S2 Oxides Ellipsometry.** Angular scan ellipsometry data and fitting model (black dashed lines) for **a,** $SiO_2$ and **b,** $TiO_2$ layers on Si substrate. n and k values of **c,** $SiO_2$ (left) and **d,** $TiO_2$ (right) extrapolated from the ellipsometric model.

Ellipsometric measurements are performed also on 35 nm of Lumogen Red F305 deposited on glass substrate; the results are reported in Fig. S3.

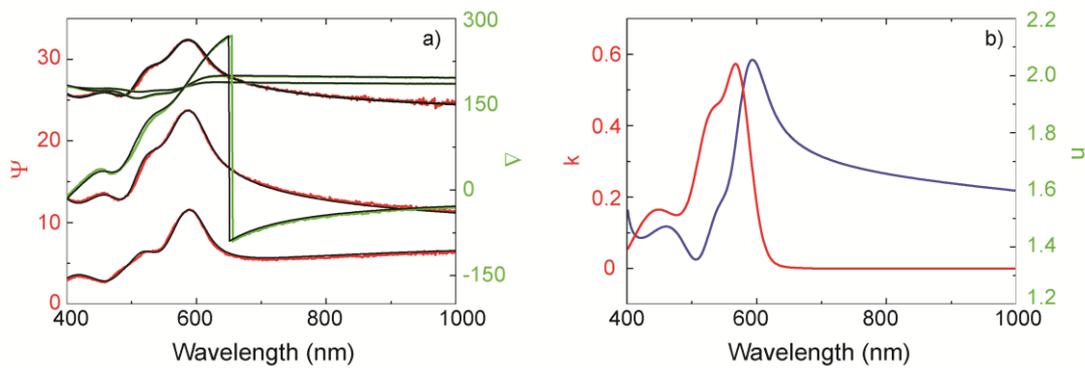

**Figure S3 Lumogen Ellipsometry. a,** Angular scan ellipsometry data and fitting model (black lines) of Lumogen Red F305. **b,** n and k values of Lumogen Red F305 extrapolated from the ellipsometric model.

The complex refractive indexes extrapolated with ellipsometric measurements are also introduced into the TMM calculation for checking the experimental dispersion map of the DBRs and the DBR/organic layer systems.

## Group Velocity

In Fig. S4 the group velocities of the BSWP vs energy are reported. In our range of analysis the group velocity exceeds 120 μm/ps and, when close to the light-line, it can reach values beyond the light velocity in the organic medium.

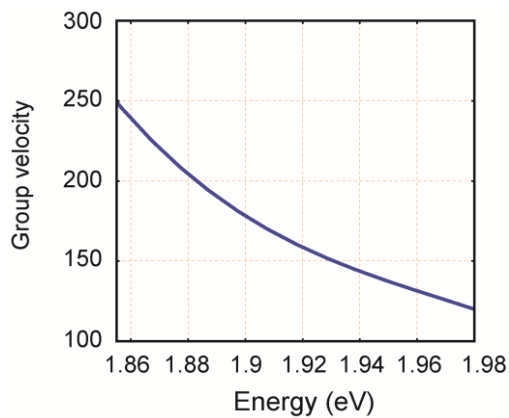

**Figure S4 Group velocity.** Group velocity obtained directly from the first derivative of the dispersion.

## Reversibility

To prove reversibility of the nonlinear effect shown in Fig. 4 of the main text, in Fig. S5 the dispersion of the BSWP before and after 50000 pulses at excitation densities of 10 mJ/cm$^2$ is plotted. As can be seen, the original dispersion (dark blue line) is recovered completely (cyan line) with no sign of degradation on the sample. On the contrary, for higher excitation densities, the dispersion changes irreversibly as shown in Fig. S6a.

We would like to stress that the reversible blueshift is exciton fraction dependent (see Fig. 4), therefore leading to a different shift for a given k vector. At higher excitonic fraction the dispersions in the linear and nonlinear regimes strongly differ, while far from the exciton resonance the two dispersions change only slightly.

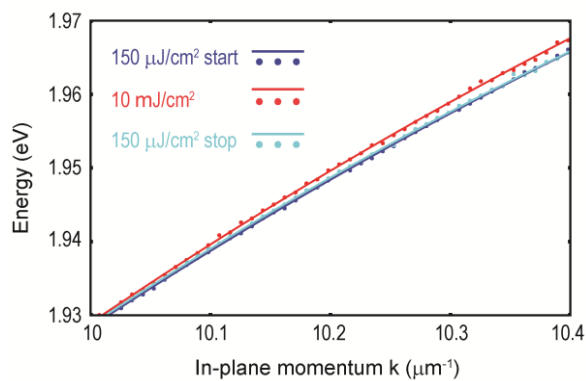

**Figure S5 Reversibilty.** BSWP dispersion measured before (blue) and after (cyan) 50000 pulses at 10 mJ/cm² (red). The dots represent the bare data of the BSWP, extracted from the reflectivity dip minimum, lines are obtained by interpolation.

We note here that this behavior can be explained only by polariton-polariton interaction. If a modification of the effective refractive index, or a melting-induced thinning of the organic layer occurs–which is indeed the case for much higher pumping powers–the opposite trend of the blueshift is observed. Figure S6 shows the effect of layer thinning (due to thermal melting) and/or quenching of the absorbers at very high pumping powers using 50000 pulses at 20 mJ/cm². The variation of the dispersions from the initial to the final measurement is consistent with an irreversible shift of the optical mode. A further proof that in this case the effect is not due to the exciton renormalization can be seen from the graph of Fig. S6b in which the difference between the initial and final state is shown. Here it is clear that the maximum effect is due to the optical mode shift rather than the exciton blueshift. Indeed, under such high power the main effect is the shift of the resonance of the optical mode and the BSWP dispersion appears blue-shifted close to the light line rather than the exciton line which, instead, remains unchanged.

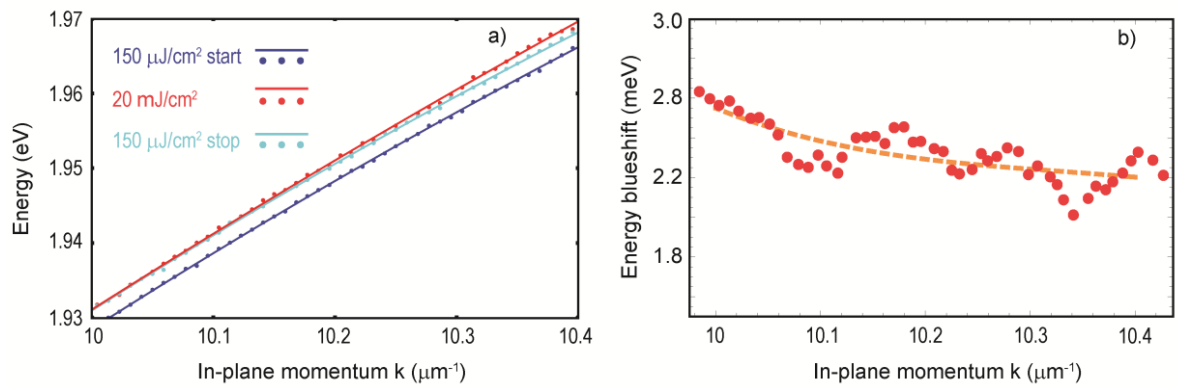

**Figure S6 Irreversibility. a,** BSWP dispersion data (dots) and interpolation (line) measured before (blue) and after (cyan) 50000 pulses at 20 mJ/cm² (red). **b,** Energy difference of the polariton dispersion before and after the high energy pulse due to the damage of the sample.